\documentclass[showkeys,12pt, preprint,preprintnumbers,nofootinbib, groupedaddress,superscriptaddress,amsmath,amssymb]{revtex4}
\usepackage{graphicx}
\usepackage{dcolumn}
\usepackage{bm}
\usepackage{amssymb}
\usepackage{amsmath}
\usepackage{epsfig}    
\usepackage{color}
\usepackage{slashed}
\usepackage{hhline}
\def\be{\begin{equation}}
\def\ee{\end{equation}}
\newcommand{\bea}{\begin{eqnarray}}
\newcommand{\eea}{\end{eqnarray}}
\newcommand{\nn}{\nonumber}

\numberwithin{equation}{section}

\begin{document}

\title{A testable radiative neutrino mass model with multi-charged particles }

\author{Kingman Cheung}
\email{cheung@phys.nthu.edu.tw}
\affiliation{Physics Division, National Center for Theoretical Sciences, 
Hsinchu, Taiwan 300}
\affiliation{Department of Physics, National Tsing Hua University, 
Hsinchu 300, Taiwan}
\affiliation{Division of Quantum Phases and Devices, School of Physics, 
Konkuk University, Seoul 143-701, Republic of Korea}

\author{Hiroshi Okada}
\email{macokada3hiroshi@cts.nthu.edu.tw}
\affiliation{Physics Division, National Center for Theoretical Sciences, 
Hsinchu, Taiwan 300}

\pacs{}
\date{\today}

\begin{abstract}
We propose a radiatively-induced neutrino mass model at one-loop level by introducing 
a pair of doubly-charged fermions and a few multi-charged bosons.  
We investigate the contributions of the model to neutrino masses,
lepton-flavor violations, muon $g-2$, oblique parameters, and collider
signals, and find a substantial fraction of the parameter space that
can satisfy all the constraints.  
Furthermore, we discuss the possibility of detecting the doubly-charged fermions at the LHC.
\end{abstract}
\maketitle

\section{Introduction}

Neutrino oscillation experiments have accumulated enough evidences that
the neutrinos do have masses.
{Massive neutrino is one of the established evidences 
beyond the standard model (SM)}. 
In order to reconcile the tiny neutrino mass to the mass of other SM fermions, 
many different mechanisms have been proposed to explain the neutrino masses.
One of the ideas that the scale of the neutrino Yukawa couplings 
should not be too different from the other Yukawa couplings --
radiatively induced neutrino mass scenario -- the neutrino is generated 
at loop level while the tree-level one is forbidden~\cite{a-zee, Cheng-Li, Pilaftsis:1991ug, Ma:2006km}.
Because of loop
suppression, small enough neutrino masses can be generated.
At the same time, it requires new fields that run inside the loop(s) 
of the neutrino-mass generating diagrams.
These new fields may be of interests to explain other phenomena,
such as dark matter, muon anomalous magnetic moment, and/or 
to give interesting signatures at the Large Hadron Collider (LHC).

In this work, we propose a simple extension of the SM
by introducing 3 generations of doubly-charged fermion pairs 
and three multi-charged bosonic fields~\cite{Aoki:2011yk}.
All of them participate in generation of neutrino mass at one-loop
level. We show that the model can explain the anomalous magnetic 
moment without conflict
constraints of the lepton-flavor violating processes and oblique
parameters. Also we discuss the possibility of detecting some of the 
new fields at the LHC.

This paper is organized as follows.
In Sec.~II, we review the model, describe several constraints, 
and show numerical results.
In Sec.~III, we discuss the collider signatures.
We conclude in Sec.~IV.

\begin{table}[t]
\begin{tabular}{|c||c||c|c|c|}
\hline\hline  
& E & ~$k^{++}$~& ~$\Phi_{\frac32}$~  & ~$\Phi_{\frac52}$ \\\hline 
$SU(2)_L$ & $\bm{1}$  & $\bm{1}$  & $\bm{2}$  & $\bm{2}$  \\\hline 
$U(1)_Y$   & $-2$   & $2$ & $\frac32$ & $\frac52$    \\\hline
\end{tabular}
\caption{\small 
Charge assignments of new fields under $SU(2)_L\times U(1)_Y$,
where all these fields are singlet under $SU(3)_C$.}
\label{tab:1}
\end{table}

\section{Model setup and Constraints}
In the model, we introduce three families of doubly charged fermions $E$,
and three types of new bosons $k^{++}$, $\Phi_{3/2}$ and  $\Phi_{5/2}$,
in addition to the SM fields, as shown in Table~\ref{tab:1}.
Under their charge assignments, 
the relevant Yukawa Lagrangian and the non-trivial terms of Higgs potential 
are given by 
\begin{align}
-\mathcal{L}_{Y}
&=
 f_{ia} \bar L_i P_R E_a \Phi_{3/2} + M_{E_a} \bar E_a E_a + \kappa_{ij} \bar e_i P_Re^c_j k^{--}
+ {g_{ia} \bar L_i P_R E^c_a \Phi^*_{5/2} }
+ {\rm h.c.},\label{Eq:lag-yukawa} \\
V
&=\left[\mu (H^T\cdot \Phi_{\frac32})k^{--} +{\rm c.c.}\right]
+\left[\mu' (H^\dag \Phi_{\frac52})k^{--} +{\rm c.c.}\right]\nn\\
&+\left[\lambda_0 (H^T\cdot \Phi_{\frac32}) (H^T\cdot \Phi_{\frac52}^*) +{\rm c.c.}\right]
+\left[\lambda_0' (\Phi_{\frac52}^\dag \Phi_{\frac32})_{\bf 3} (H^T H)_{\bf 3} +{\rm c.c.}\right],
\label{Eq:lag-pot}
\end{align}
where $H$ is the SM Higgs field that develops a nonzero vacuum 
expectation value (VEV), which is symbolized by 
$\langle  H \rangle\equiv v/\sqrt2$, and $(i,a)=1-3$ are generation indices.
The $f$ and $g$ terms contribute to the active neutrino masses, 
while the $\kappa$ term does not contribute to the neutrino sector 
but plays a role {of mediating the decays of the new particles into the SM particles.}
{In this work, all the coefficients are chosen to be real and positive for simplicity. }

We parameterize the scalar fields as
\begin{align}
&\Phi_{\frac32} =\left[
\begin{array}{c}
\phi^{++}_{3/2} \\
\phi^{+}_{3/2}
\end{array}\right],
\quad 
\Phi_{\frac52} =\left[
\begin{array}{c}
\phi^{+++}_{5/2} \\
\phi^{++}_{5/2}
\end{array}\right],
\label{eq:component}
\end{align}
where the lower index in each component represents the 
{hypercharge of the field.}
Due to the $\mu^{(')}$ and $\lambda_0^{(')}$ terms in 
Eq.~(\ref{Eq:lag-pot}), the three doubly-charged bosons in basis of 
$(k^{++},\phi^{++}_{3/2},\phi^{++}_{5/2})$ fully mix {with} 
one another.
%
The mixing matrix and mass eigenstates are defined as follows: 
\begin{align}
&\left[\begin{array}{c} k^{++} \\ \phi^{++}_{3/2} \\ \phi^{++}_{5/2} \end{array}\right] = 
\sum_{a=1-3}O_{ia} H_{a}^{++},\quad
O\equiv 
\left[\begin{array}{ccc} 1 & 0 & 0 \\
0 & c_{23} & s_{23} \\
0 &  -s_{23} & c_{23}   \end{array}\right]
\left[\begin{array}{ccc} 
 c_{13} & 0 & s_{13} \\
0 & 1 & 0 \\
  -s_{13} & 0 & c_{13}   \end{array}\right]
\left[\begin{array}{ccc} 
 c_{12} & s_{12} & 0  \\
  -s_{12}  & c_{12}  & 0 \\
  0 & 1 & 0 \\
   \end{array}\right],  
\end{align}
therefore  one can rewrite the Lagrangian in terms of  
the mass eigenstate as follows:
\begin{align}
 k^{++} = \sum_{a=1-3}O_{1a} H_{a}^{++},\quad
 \phi^{++}_{3/2}= \sum_{a=1-3}O_{2a} H_{a}^{++},\quad
 \phi^{++}_{5/2}= \sum_{a=1-3}O_{3a} H_{a}^{++}.\label{eq:basis}
\end{align}

\begin{figure}[tb]
\begin{center}
\includegraphics[width=80mm]{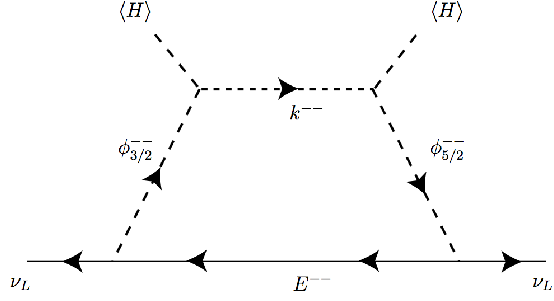}
\caption{One-loop diagram for generating the neutrino mass matrix.}
\label{fig:neutrino}
\end{center}
\end{figure}

\subsection{Neutrino mixing}
The active neutrino mass matrix $M_\nu$ is given at one-loop level
via doubly-charged particles in Fig.~\ref{fig:neutrino}, and its formula is given by
\begin{align}
&(M_{\nu})_{ij}
=\frac{2}{(4\pi)^2}\sum_{a=1}^3
{ f_{ia} M_{E_a} g_{aj}^T }
\left[\zeta_{12} F_I({E_a},H_1^{++},H_2^{++}) 
-\zeta_{13} F_I({E_a},H_1^{++},H_3^{++}) 
+\zeta_{23} F_I({E_a},H_2^{++},H_3^{++}) 
\right] 
\nn\\
&
\hspace{3cm}+ (f\leftrightarrow g)  \equiv  f_{ia} R_a g_{aj}^T + g_{ia} R_a f_{aj}^T
,\\
&F_I(a,b,c)
=\frac{m_a^2 m_b^2\ln\left(\frac{m_a}{m_b}\right)+m_a^2 m_c^2\ln\left(\frac{m_a}{m_c}\right)+m_b^2 m_c^2\ln\left(\frac{m_b}{m_c}\right)}
{(m_a^2-m_b^2)(m_a^2-m_c^2) },\\
&R_a
=\frac{2 M_{E_a}}{(4\pi)^2}
\left[\zeta_{12} F_I({E_a}, H_1^{++}, H_2^{++}) 
-\zeta_{13} F_I({E_a}, H_1^{++}, H_3^{++}) 
+\zeta_{23} F_I({E_a}, H_2^{++}, H_3^{++}) 
\right] ,
\end{align}
where $\zeta_{12}\equiv s_{12}^2 s_{13}^2s_{23}c_{23}+2 c_{12}s_{12}s_{13}s_{23}^2-c_{12}s_{12}s_{13}+s_{12}^2s_{23}c_{23}$,
$\zeta_{13}\equiv s_{13}^2 s_{23} c_{23}$, and $\zeta_{23}\equiv  s_{23} c_{23}$.
$M_\nu$ is diagonalized by the neutrino mixing matrix $V_{\rm MNS}$ as 
$M_\nu  = V_{\rm MNS} D_\nu V_{\rm MNS}^T$ with $D_\nu\equiv 
{{\rm diag}}(m_{\nu_1},m_{\nu_2},m_{\nu_3})$. 
Then one can parameterize the Yukawa coupling in terms of an 
arbitrary antisymmetric matrix $A$ with complex values (i.e. $(A+A^T=0))$ {with mass scale},
as follows~\cite{Okada:2015vwh, Cheung:2016fjo}:
\begin{align}
& f=\frac12 [V_{\rm MNS} D_\nu V_{\rm MNS}^T+A] (g^T)^{-1} R^{-1},
\quad
 g=\frac12 [V_{\rm MNS} D_\nu V_{\rm MNS}^T+A]^T (f^T)^{-1} R^{-1}.
\end{align}
In the numerical analysis, we use {\it the latter} relation for
convenience, and we use the
data in the global analysis~\cite{Forero:2014bxa}.
{Notice here that the mass scale of $A$ should be rather tiny so that 
 $A$ can be the relevant mass parameter to make a significant contribution to the observed neutrino oscillation data.   }

\begin{figure}[tb]
\begin{center}
\includegraphics[width=80mm]{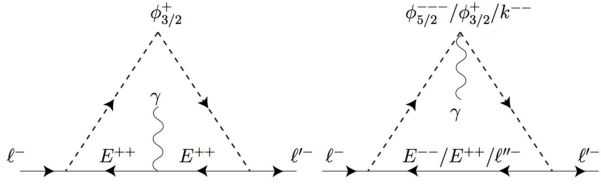}
\caption{One-loop diagrams for generating the lepton flavor violations, where $(g-2)_\mu$ is also induced from the same figure by sending $\ell(\ell')\to\mu$.}
\label{fig:lfvs}
\end{center}
\end{figure}

\subsection{Lepton flavor violations (LFVs) and muon $g-2$}
\label{lfv-lu}
The Yukawa terms of ($f,g,\kappa$) in the Lagrangian 
contribute to the lepton-flavor violating processes 
$\ell\to\ell'\gamma$ at one-loop level as shown in Fig.~\ref{fig:lfvs}.
Here the left side {of Fig.~\ref{fig:lfvs} } arises from the term 
$f$ mediated by $\phi^+_{3/2}$ and $E^{++}$,
while the right side arises from the terms $g/f/\kappa$ that 
respectively correspond to $\phi^{---}_{5/2}/\phi^+_{3/2}/k^{--}$ 
and $E^{--}/E^{++}/\ell''^{-}$.
 %
The branching ratio is given by
\begin{align}
B(\ell_i\to\ell_j \gamma)
\approx 
\frac{48\pi^3 \alpha_{\rm em}}{{\rm G_F^2} } C_{ij} |{\cal M}_{ij}|^2,
\end{align}
where $G_{\rm F}\approx1.16\times 10^{-5}$ GeV$^{-2}$ is the Fermi constant, $\alpha_{\rm em}\approx1/137$ is the fine structure constant, 
$C_{21}=1$, $C_{31}=0.1784$, and $C_{32}=0.1736$.
${\cal M}(= {\cal M}_f+{\cal M}_g+{\cal M}_\kappa)$ is formulated 
as
\begin{align}
&({\cal M}_f)_{ij}
\approx
-\sum_{a=1-3}\frac{f_{ja} f^\dag_{ai} }{(4\pi)^2} \left[{F_{lfv}({E_a},{\phi_{3/2}^{+}}) +2 F_{lfv}}({\phi_{3/2}^{+}},{E_a})\right], \label{eq:g2-f}\\
&({\cal M}_g)_{ij}
\approx
\sum_{a=1-3}\frac{g_{ja} g^\dag_{ai} }{(4\pi)^2} \left[3{F_{lfv}({E_a},{\phi_{5/2}^{+++}}) +2 F_{lfv}}({\phi_{5/2}^{+++}},{E_a})\right],\\
&({\cal M}_\kappa)_{ij}
\approx
\sum_{a,\alpha=1-3}\frac{\kappa_{ja} \kappa^\dag_{ai} |O_{1\alpha}|^2 }{3(4\pi)^2 m_{H_\alpha}^2},\\
&{F_{lvs}(a,b)=\frac{2 m_a^6+3m_a^4m_b^2-6m_a^2m_b^4+m_b^6+12m_a^4 m_b^2\ln\left[\frac{m_b}{m_a}\right]}
{12(m_a^2-m_b^2)^4}} \;.
\end{align} 
The current experimental upper bounds are given 
by~\cite{TheMEG:2016wtm, Adam:2013mnn}
  \begin{align}
  B(\mu\rightarrow e\gamma) &\leq4.2\times10^{-13},\quad 
  B(\tau\rightarrow \mu\gamma)\leq4.4\times10^{-8}, \quad  
  B(\tau\rightarrow e\gamma) \leq3.3\times10^{-8}~.
 \label{expLFV}
 \end{align}

{\it The muon anomalous magnetic moment $(g-2)_\mu$}:
It is known that discrepancy of experimental value and the SM prediction is given by~\cite{Hagiwara:2011af}
\begin{align}
\Delta a_\mu = (26.1\pm8.0)\times10^{-10}.\label{eq:damu}
\end{align}
We have nonvanishing $(g-2)_\mu$, and its formula is found via ${\cal M}$ in LFVs as
\begin{align}
\Delta a_\mu\approx -m_\mu^2 {\cal M}_{22}.
\end{align}
Here the $f$ term contribution provides the positive value of $(g-2)_\mu$ that corresponds to Fig.~\ref{fig:lfvs} with $\phi^+$ and $E^{++}$ mediators inside the loop,
while the other terms $g$ and $\kappa$  give the negative values of $(g-2)_\mu$.~\footnote{{The sign of $(g-2)_\mu$, which is induced at
    one-loop level, generally depends on sign of the electric charge and
    the direction of momentum of the particle that emits the photon inside
    the loop.  For example, when a fermion (boson) with
    negative (positive) electric charge propagates in the same direction
    as the outgoing  muon, one finds positive values for
    $(g-2)_\mu$.  In the opposite case, one obtains negative
    $(g-2)_\mu$. Through this aspect, one can straightforwardly
    understand the sign of $(g-2)_\mu$ without any computations, and
    our sign shows the direct consequence of this insight.}}
In order to achieve the agreement with the experimental value, 
one has to enhance $f$ term compared to the $g$ and $\kappa$ term.
However the $\kappa$ term gives another LFV with three body 
decay $\ell_i\to\ell_j\ell_k\bar\ell_\ell$ at tree level and it gives more stringent constraints as shown in Table I of 
Ref.~\cite{Herrero-Garcia:2014hfa}.
Thus we can expect this term to be negligible in $(g-2)_\mu$.

\subsection{Oblique parameters} 
\label{subseq.st}
In order to estimate the testability via collider physics, 
we have to consider the oblique parameters that restrict the 
mass hierarchy between each of the components in
$\Phi_{\frac32}$ and $\Phi_{\frac52}$.

Here we focus on the new physics contributions to $S$ and $T$ 
parameters in the case $U=0$.
Then  
$\Delta S$ and $\Delta T$ are defined as
\begin{align}
\Delta S&={16\pi} \frac{d}{dq^2}[\Pi_{33}(q^2)-\Pi_{3Q}(q^2)]|_{q^2\to0},\quad
\Delta T=\frac{16\pi}{s_{W}^2 m_Z^2}[\Pi_{\pm}(0)-\Pi_{33}(0)],
\end{align}
where $s_{W}^2\approx0.23$ is the Weinberg angle and $m_Z$ is the $Z$ 
boson mass. 
The loop factors $\Pi_{33,3Q,\pm}(q^2)$ are calculated from the one-loop 
vacuum-polarization diagrams for $Z$ and $W^\pm$ bosons, which are respectively given by
{\begin{align}
\Pi_{33}&=
\frac{1}{2(4\pi)^2}\left[
{G(q^2,m^2_{\phi^+_{3/2}},m^2_{\phi^+_{3/2}})}
+{(|O_{2\alpha}|^2+|O_{3\alpha}|^2)}
\left[G(q^2,m^2_{H^{++}_\alpha},m^2_{H^{++}_\alpha}) - H(m_{H^{++}_\alpha}^2)\right]
\right.\nn\\
&\left.
+{G(q^2,m^2_{\phi^{+++}_{5/2}},m^2_{\phi^{+++}_{5/2}})}
-{H(m^2_{\phi^+_{3/2}})}
-{H(m^2_{\phi^{+++}_{5/2}})}
\right],\\
\Pi_{3Q}&=
\frac{1}{(4\pi)^2}\left[
-{G(q^2,m^2_{\phi^+_{3/2}},m^2_{\phi^+_{3/2}})}
+2{(|O_{2\alpha}|^2-|O_{3\alpha}|^2)}
\left[G(q^2,m^2_{H^{++}_\alpha},m^2_{H^{++}_\alpha}) - H(m_{H^{++}_\alpha}^2)\right]\right.\nn\\
&\left.
+3{G(q^2,m^2_{\phi^{+++}_{5/2}},m^2_{\phi^{+++}_{5/2}})}
+ {H(m^2_{\phi^+_{3/2}})}
-3{H(m^2_{\phi^{+++}_{5/2}})}
\right],\\
\Pi_{\pm}&=
\frac{1}{2(4\pi)^2}\left[
2{|O_{2\alpha}|^2}
G(q^2,m^2_{\phi_{3/2}},m^2_{H^{++}_\alpha}) +
2{|O_{3\alpha}|^2}
G(q^2,m^2_{\phi_{5/2}},m^2_{H^{++}_\alpha}) \right.\nn\\
&\left.
- {(|O_{2\alpha}|^2+|O_{3\alpha}|^2)}H(m_{H^{++}_\alpha}^2)
-{H(m^2_{\phi^+_{3/2}})}
-{H(m^2_{\phi^{+++}_{5/2}})}\right].
\end{align}
}
The experimental bounds are given by \cite{pdg}
\begin{align}
(0.05 - 0.09) \le \Delta S \le (0.05 + 0.09), \quad 
(0.08 - 0.07) \le \Delta T \le (0.08 + 0.07),
 \end{align}
 and new contributions should be within these ranges.

\subsection{Numerical analysis \label{sec:numerical}}

In the numerical analysis,
we prepare $2\times10^6$ random sampling points for the relevant input 
parameters in the following ranges:
\begin{align}
& s_{12,23,13}\in  [-0.1\,, 0.1\,],\quad (A_{12},A_{13},A_{23}) \in  \pm {[10^{-18},10^{-8}]\ \text{TeV}}
,\nn\\
& (f_{11},f_{12},f_{13}) \in  \pm [10^{-10},10^{-5}], \quad (f_{21},f_{22},f_{23}) \in  \pm [1,4\pi], \quad
 \quad (f_{31},f_{32},f_{33}) \in  \pm [10^{-3},1],
\nn\\
& m_{H_1^{++}} \in [0.1\,, 2\,]\ \text{TeV},\quad m_{H_2^{++}} \in [m_{H_1^{++}}\,, 2\,]\ \text{TeV},\quad m_{H_3^{++}} \in [m_{H_2^{++}}\,, 2\,]\ \text{TeV},\nn\\
& m_{\phi^+_{3/2}} \in [m_{H_2^{++}} \pm 0.1]\ \text{TeV},\quad 
m_{\phi^{+++}_{5/2}} \in [m_{H_3^{++}} \pm 0.1]\ \text{TeV},\nn\\
& M_1 \in [m_{\phi^{+++}_{5/2}} \,, 2\,]\ \text{TeV},\quad M_2 \in [M_1\,, 2\,]\ \text{TeV},\quad M_3 \in [M_2\,, 2\,]\ \text{TeV},
\label{range_scanning}
\end{align}
and we find {650} allowed points that satisfy 
neutrino oscillation data, LFVs, oblique parameters, 
and  observed $(g-2)_\mu$: $\Delta a_\mu=(26.1\pm8.0)\times10^{-10}$ 
in Eq.~(\ref{eq:damu}).
Here we take rather large Yukawa couplings $f_{21},f_{22},f_{23}$ in order to 
obtain sizable $(g-2)_\mu$.
{On the other hand,}
$f_{11},f_{12},f_{13}$ have to be tiny in order to satisfy the stringent 
constraint of $\mu\to e\gamma$ process, which is proportional to
 $f_{11}f_{21}+f_{12}f_{22}+f_{13}f_{23}$, while 
$f_{31},f_{32},f_{33}$ are taken to be of typical scale to satisfy 
the other LFVs.

\begin{figure}[t]
\includegraphics[width=80mm]{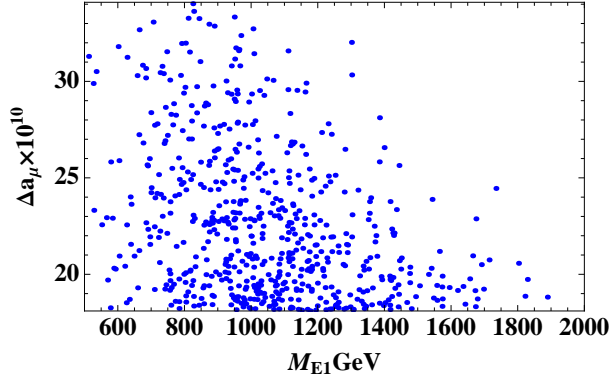}
\caption{
Scatter plot of $\Delta a_\mu \times 10^{10}$ versus $M_{E_1}$ GeV 
{that satisfy all the constraints mentioned in the text}.}
\label{fg:M1-damu}
\end{figure}

In Fig.~\ref{fg:M1-damu}, we show the {
scatter plot in the plane of $M_{E_1}$ and $\Delta a_\mu$ that 
satisfy all the constraints as discussed above.
We observe that the whole mass range of $E_1$ that we have taken 
can give $\Delta a_\mu$ to be within $(26.1\pm8.0)\times10^{-10}$.
Also, the smaller the mass $M_{E_1}$ the larger the value of $\Delta a_\mu$ 
will be, as expected by the formula in Eq.~(\ref{eq:g2-f}).}

\begin{figure}[t]
\includegraphics[width=80mm]{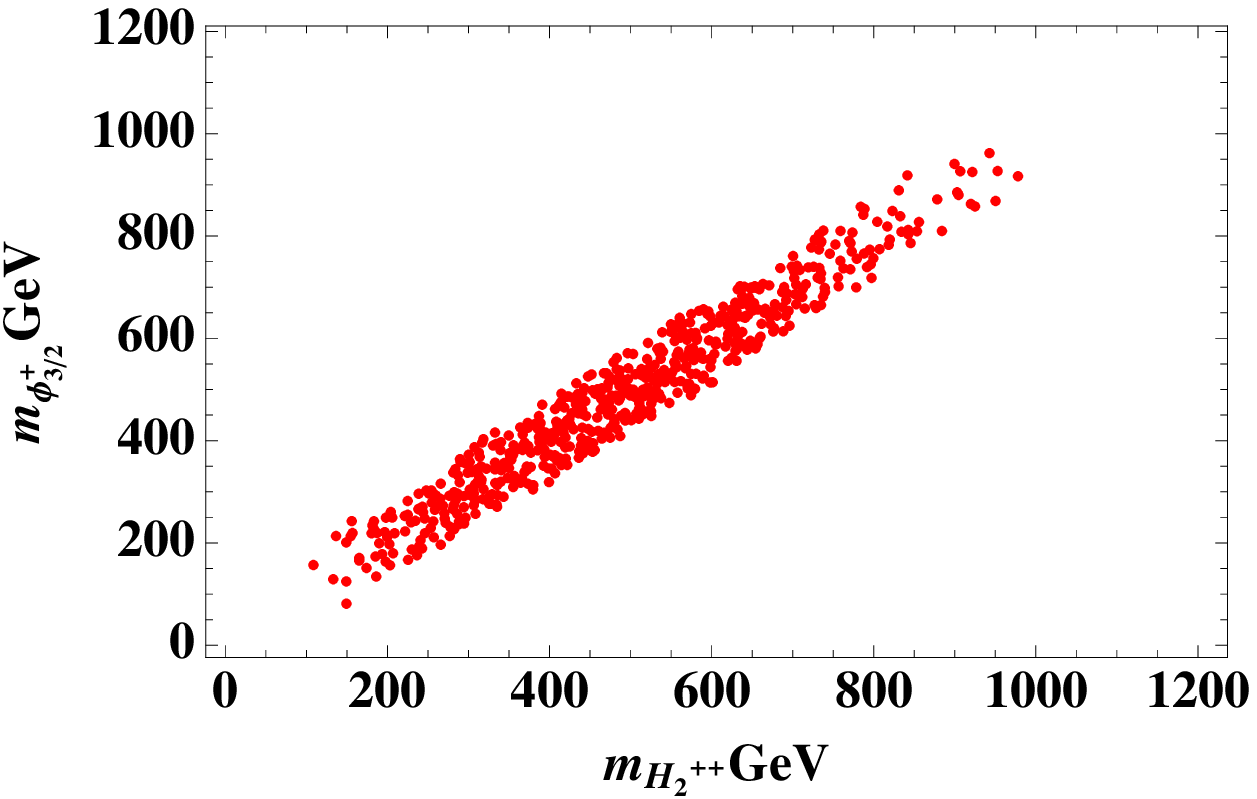}
\includegraphics[width=80mm]{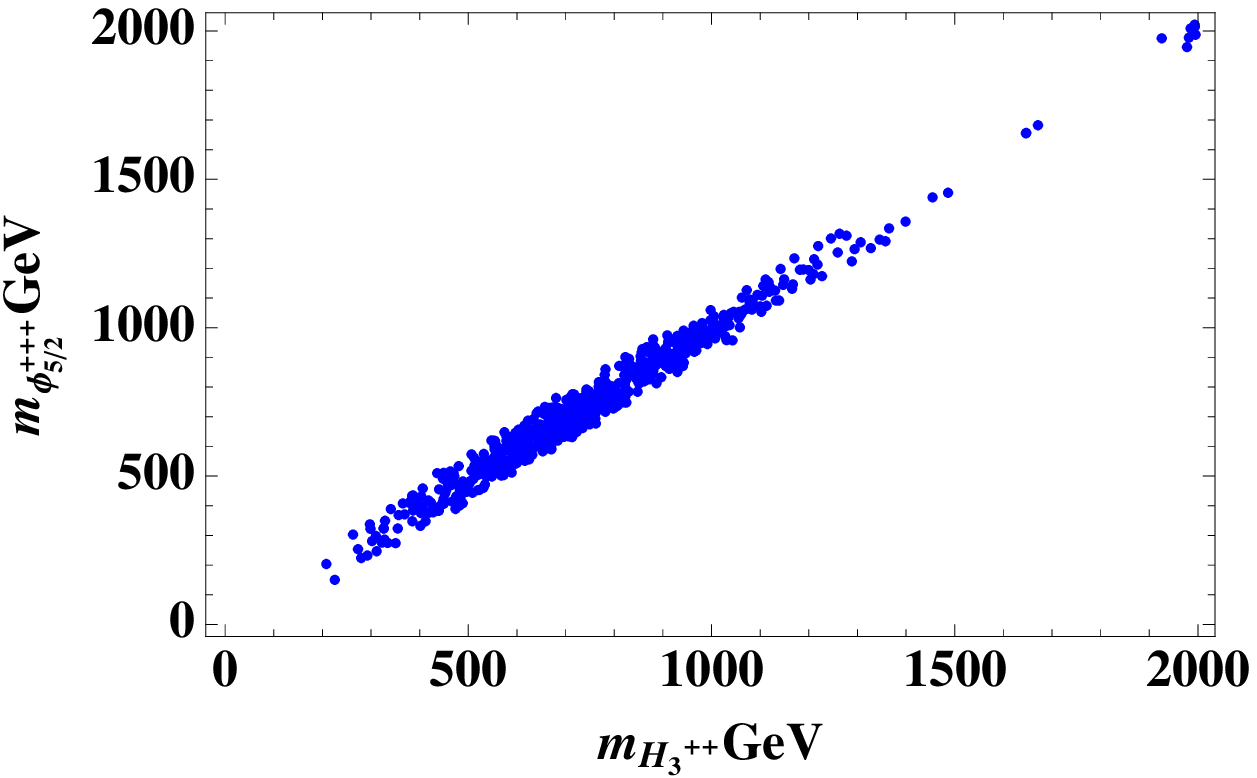}
\caption{
Scatter plots of $m_{\phi^+_{3/2}}$ versus $m_{H_2^{++}}$ GeV with red points 
on the left side,
and 
$m_{\phi^{+++}_{5/2}}$ versus $m_{H_3^{++}}$ GeV  with blue points on the right side. 
Notice here that only these two pairs have strong correlations 
due to the oblique parameters.
}
\label{fg:bm}
\end{figure}

In Fig.~\ref{fg:bm}, we show two characteristic {
correlations} among the masses of the charged bosons.
Theses {correlations} suggest that the masses 
between $m_{H_2^{++}}\, (m_{H_3^{++}})$ and 
$m_{\phi_{3/2}^+}\, (m_{\phi_{5/2}^{+++}} )$ 
{are almost degenerate.}
Since we {have taken small mixings} among the doubly-charged 
bosons in the input parameters, such mass degeneracy naturally
occurs in each of the isospin doublets. This is a consequence of 
the constraint from the oblique parameters as discussed in Sec.~\ref{subseq.st}.
The second {feature} is that the mass range of 
$m_{\phi_{3/2}^+}$ is restricted to be {less than} 1 TeV, 
even though we have scanned it up to 2 TeV as input parameters. 
This mainly comes from the experimental value of $\Delta a_\mu$.
Moreover, the value of the loop function in (\ref{eq:g2-f}) decreases 
when the mass of $\phi_{3/2}^+$ increases.
Then $m_{H_2^{++}}$ is also restricted to be in the 
same {range} as 
$m_{\phi_{3/2}^+}$ by the consequence of oblique parameters again.



\section{Collider Signals}

We first consider the Drell-Yan (DY) production of $E \overline{E}$
via $\gamma,\, Z$ exchanges. The interactions can be obtained from the
kinetic term of the fermion $E$. Since $E$ is a singlet, the interactions
with $\gamma$ and $Z$ are given by
\[
  {\cal L}  = - e \overline{E} \gamma^\mu Q_E E\, A_\mu 
              + \frac{g s_W^2 }{c_W} \overline{E} \gamma^\mu Q_E E\, Z_\mu \;,
\]
where $s_W$ and $c_W$ are respectively the sine and cosine of the Weinberg
angle, and $Q_E$ is the electric charge of the fermion $E$ with $Q_E = -2$
in our model. 

\begin{figure}[th!]
\begin{center}
\includegraphics[width=100mm]{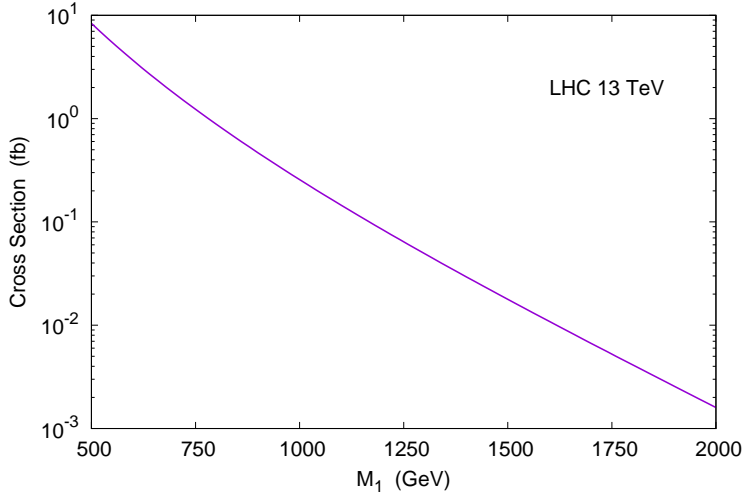}
\vspace{10mm}
\caption{
Drell-Yan Production cross section for $pp \to E^{++}_1 E^{--}_1$ at the LHC-13.}
\label{fg:cross}
\end{center}
\end{figure}

The square of the scattering amplitude, summed over spins, for 
$q (p_1) \bar q (p_2) \to E (k_1) \overline{E} (k_2)$ can be written as
\begin{eqnarray}
\sum |{\cal M}|^2 &=& 4 e^4 Q_E^2 \left[ 
   \left(\hat u - {M}^2_E \right )^2 +    \left(\hat t - {M}^2_E \right )^2 
  + 2 \hat s {M}_E^2 \right ] \nonumber \\
& \times & \left\{
        \left| \frac{Q_q}{\hat s} - \frac{g_L^q}{ c_W^2 } \frac{1}{\hat s -m_Z^2}
             \right|^2
+
        \left| \frac{Q_q}{\hat s} - \frac{g_R^q}{ c_W^2 } \frac{1}{\hat s -m_Z^2}
             \right|^2 
  \right \} \;,
\end{eqnarray}
where $\hat s, \, \hat t, \, \hat u$ are the usual Mandelstam variables
for the subprocess, and 
{ $g_{L}^q = T_{3q} - s_W^2 Q_q$ and $g_{R}^q = - s_W^2 Q_q$}
are the chiral couplings of quarks to the $Z$ boson.
The subprocess differential cross section is given by
\begin{eqnarray}
\frac{d \hat \sigma}{ d\cos \hat \theta} &=& \frac{\beta  e^4 Q_E^2}{96\pi}
 \left[ 
   \left(\hat u - {M}^2_E \right )^2 +    \left(\hat t - {M}^2_E \right )^2 
  + 2 \hat s m_E^2 \right ] \nonumber \\
& \times & \left\{
        \left| \frac{Q_q}{\hat s} - \frac{g_L^q}{ c_W^2 } \frac{1}{\hat s -m_Z^2}
             \right|^2
+
        \left| \frac{Q_q}{\hat s} - \frac{g_R^q}{ c_W^2 } \frac{1}{\hat s -m_Z^2}
             \right|^2 
  \right \} \;,
\end{eqnarray}
where $\beta = \sqrt{ 1- 4 {M}_E^2 / \hat s}$,
{and
where $T_{3q}$ is the third component of the isospin of $q$.}
This subprocess cross section is then folded with parton distribution 
functions to obtain the scattering cross section at the $pp$ collision level.
The $K$ factor for the production cross sections is expected to be similar
to the conventional DY process, which is approximately $K \simeq 1.3$ at 
the LHC energies.
{The production cross sections for $pp \to E^{++}_1 E^{--}_1$ at $\sqrt{s}=13$ TeV
LHC are shown in Fig.~\ref{fg:cross}. For $M_{E_1} \approx 1$ TeV the
cross section is about $0.2$ fb.  
}

We proceed to estimate the decay partial widths of the fermion $E^{--}_1$,
which is presumed to the lightest among $E^{--}_{1,2,3}$.
The decay channels of $E^{--}_1$ can proceed via the following 
interactions
\begin{eqnarray}
{\cal L} &=& f_{i1} \left[ \overline{\nu_i} P_R E_1^{--} \,
  \left( O_{21} H_1^{++}  + O_{22} H_2^{++}  +O_{23} H_3^{++} \right )
    + \bar e_i P_R E^{--}_1 \, \phi_{3/2}^+ \right  ]
 \nonumber \\
\label{decayE}
&+& g_{i1} \left [ \overline{\nu_i} P_R E^c_{1} \, 
  \left( O_{31} H_1^{--}  + O_{32} H_2^{--}  +O_{33} H_3^{--} \right )
    + \bar e_i P_R E^c_1 \, \phi_{5/2}^{---} \right  ] \;.
\end{eqnarray}
We shall take the approximation that the diagonalizing matrix $O$ is nearly 
diagonal, such that $O_{11}, O_{22}, O_{33} \approx 1$.  In such a case,
$H_1^{++} \approx k^{++}$, 
$H_2^{++} \approx \phi_{3/2}^{++}$ and $H_3^{++} \approx \phi_{5/2}^{++}$.
We also take the simplification that the masses of each components in the
doublet are similar, i.e.,  ${m}_{\phi_{3/2}^{+} } \approx  {m}_{\phi_{3/2}^{++}}$ and
${m}_{\phi_{5/2}^{++} } \approx  {m}_{\phi_{5/2}^{+++}}$.

We compute the partial width of $E^{--} \to e_i \phi_{3/2}^-$ and obtain
\begin{equation}
 \Gamma( E^{--}_1 \to e_i \phi_{3/2}^- ) = \frac{|f_{i1}|^2 }{64\pi} M_{E_1}
\left(  1 - \frac{{m}_{\phi_{3/2}}^2 }{ M_{E_1}} \right )
\end{equation}
which is the same as $ \Gamma( E^{--}_1 \to \nu_i H_2^{--} ) $, in
which $H_2^{--}$ is mostly $\phi_{3/2}^{--}$.
Summing over all lepton and neutrino channels with $i=1,2,3$ as well as
the contributions from the $f_{i1}$ and $g_{i1}$ terms, we obtain
the total decay width of $E_1^{--}$
\begin{equation}
\label{widthE}
\Gamma(E_1^{--}) = \frac{M_{E_1}}{32\pi}  \left \{
\left(  1 - \frac{{m}_{\phi_{3/2}}^2 }{ M_{E_1}} \right ) \, 
     \sum_{i=1}^{3}  |f_{i1}|^2 
+ 
\left(  1 - \frac{{m}_{\phi_{5/2}}^2 }{ M_{E_1}} \right ) \, 
     \sum_{i=1}^3  |g_{i1}|^2 
\right \}
\end{equation}

Next, we compute the subsequent decays of $H_i^{--} \to e^-_k e^-_l$ (where
$k,l$ are flavors) and $\phi_{3/2}^- \to H_i^{--} W^+$:
\begin{eqnarray}
\Gamma(H_i^{--} \to e^-_k e^-_l ) &=& \frac{\kappa_{kl}^2  |O_{1i}|^2 }{16\pi} \,
    {m}_{H_i} \label{width1} \\
\Gamma( \phi_{3/2}^- \to H_i^{--} W^+ ) &=&  \frac{|O_{2i}|^2}{ 128\pi}
 \frac{{m}_{\phi_{3/2}}^3} { m_W^2} \lambda^{3/2} 
  \left( 1,\,  \frac{m_W^2}{ {m}_{\phi_{3/2}}^2 }, \, 
               \frac{m_{H_i}^2}{ {m}_{\phi_{3/2}}^2 } \right ) \label{width2}
\end{eqnarray}
where the function $\lambda(x,y,z) = (x^2 + y^2 + z^2 -2xy -2yz - 2zx)$
and if the mass difference ${m}_{\phi_{3/2}} - {m}_{H_i} < m_W$ then the latter
decay would proceed via a virtual $W$ boson. Here the parameter $\kappa_{kl}$ 
can be chosen arbitrarily so as to decay the charged boson $k^{--}$ to ensure
no stable charged particles left in the Universe. 
Therefore, each singlet fermion $E^{--}$ so produced can decay 
into 2 charged leptons or 4 charged leptons plus missing energies.
In DY production of a pair of singlet fermions $E^{--} E^{++}$, the final
state consists of 4 or 8 charged leptons plus missing energies, 
which is extremely spectacular in hadron colliders. 

Similarly, the singlet fermion $E^{--}_1$ can decay into the $\phi_{5/2}$ doublet
via the second term in the Lagrangian (\ref{decayE}), including
$E^{--} \to \phi_{5/2}^{---} \bar e_i$ and 
$E^{--} \to H_3^{--} \bar \nu_i$. These partial widths have already been 
included in Eq.~(\ref{widthE}).  The decay pattern of the components in
the $\phi_{5/2}$ doublet is
\begin{eqnarray}
&&\left (\phi_{5/2}^{--} \approx H_3^{--} \right ) \to e^-_k e^-_l \nonumber \\
&& \phi_{5/2}^{---} \to H_i^{--} W^- \nonumber \;,
\end{eqnarray}
of which their decay widths can be obtained from Eqs.~(\ref{width1})
and (\ref{width2}) by replacing ${m}_{\phi_{3/2}} \to {m}_{\phi_{5/2}}$.

Naively, since $g_{i1} \ll f_{i1}$ due to lepton-number violation, we 
expect $E^{--} \to e_i \phi^-_{3/2},\; \nu_i \phi_{3/2}^{--}$ dominantly.
Therefore, the branching ratio for $E^{--} \to 2 e_i + \not\!{E}$ is about
$1/2$, for $E^{--} \to 4 e_i + \not\!{E}$ is about $1/6$ (including
$e_i = e,\mu,\tau$).
Now we can estimate the event rates at the 13 TeV LHC with 
a luminosity of 3000 fb$^{-1}$ (HL-LHC). We have about 
$0.2 \times 3000 \times (1/2)^2 = 150$ events for $4e_i$ final state,
$0.2 \times 3000 \times 1/2 \times 1/6 \times 2 = 100$ events 
for $6e_i$ final state, and
$0.2 \times 3000 \times 1/6 \times 1/6 \simeq  17 $ events 
for $8e_i$ final state.


\section{Conclusions}

In this work, we have proposed a simple extension of the SM 
by introducing 3 generations of doubly-charged fermion pairs 
and three multi-charged bosonic fields.
We have investigated the contributions of the model to neutrino mass,
lepton-flavor violations, muon $g-2$, oblique parameters, and collider
signals, and found a substantial fraction of the parameter space that
can satisfy all the constraints.  
%

The design of the $\kappa$ term in the Lagrangian is to make sure that
all new charged particles will decay into SM particles so that no
stable charged particles were left in the Universe.  Because of this
$\kappa$ term the new charged particles will decay into charged
leptons in collider experiments, thus giving rise to spectacular
signatures. Pair production of $E^{++}_1 E^{--}_1$ can give $4e_i, 6
e_i$, or $8e_i$ plus missing energies in the final state.  The event
rates are $17-150$ for an integrated luminosity of 3000 fb$^{-1}$.

\section*{Acknowledgments}
This work was supported by the Ministry of Science and Technology
of Taiwan under Grants No. MOST-105-2112-M-007-028-MY3.


\end{document}